\documentclass[aps,prd,eqsecnum]{revtex4}
\usepackage{latexsym}
\usepackage{bm}

\begin{document}
\input epsf.sty


%
%
\leftline{\epsfbox{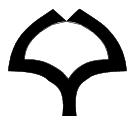}}
\vspace{-10.0mm}
{\baselineskip-4pt
\font\yitp=cmmib10 scaled\magstep2
\font\elevenmib=cmmib10 scaled\magstep1  \skewchar\elevenmib='177
\leftline{\baselineskip20pt
\hspace{12mm} 
\vbox to0pt
   { {\yitp\hbox{Osaka \hspace{1.5mm} University} }
     {\large\sl\hbox{{Theoretical Astrophysics}} }\vss}}

%
%
{\baselineskip0pt
\rightline{\large\baselineskip14pt\rm\vbox
        to20pt{\hbox{August 2002}
               \hbox{OU-TAP-xxx}
               \hbox{KUNS-1800}
\vss}}
}
\vskip15mm

\begin{center}

{\Large\bf Gauge Problem in the Gravitational Self-Force I.}

\vspace{3mm}

{\large \it --- Harmonic Gauge Approach
in the Schwarzschild Background ---}

\bigskip

Norichika Sago$^{1,2}$, Hiroyuki Nakano$^1$ and Misao Sasaki$^1$

\smallskip

$^1${\em Department of Earth and Space Science,~Graduate School of
 Science,~Osaka University,\\ Toyonaka, Osaka 560-0043, Japan
}

\smallskip
$^2${\em Department of Physics,~Graduate School of Science,
Kyoto University,\\ Kyoto 606-8502,~Japan}
\\
\smallskip

\medskip

\today

\end{center}

\bigskip

\begin{abstract}
\end{abstract}

The metric perturbation induced by a particle in the Schwarzschild
background is usually calculated in the Regge-Wheeler (RW) gauge,
whereas the gravitational self-force is known to be given by
the tail part of the metric perturbation in the harmonic gauge.
Thus, to identify the gravitational self-force correctly in a
specified gauge, it is necessary to find out a gauge transformation
that connects these two gauges. This is called the gauge problem.
As a direct approach to solve the gauge problem,
we formulate a method to calculate the metric perturbation
in the harmonic gauge on the Schwarzshild backgound.
We apply the Fourier-harmonic expansion to the metric perturbation
and reduce the problem to the gauge transformation of the
Fourier-harmonic coefficients (radial functions)
from the RW gauge to the harmonic gauge.
We derive a set of decoupled radial equations for
the gauge transformation.
These equations are found to have a simple second-order form
for the odd parity part and the forms of spin $s=0$ and $1$
Teukolsky equations for the even parity part.
As a by-product, we correct typos
in Zerilli's paper and present a set of corrected equations
in Appendix \ref{app:RWZ}.



\section{Introduction}
A compact object of solar-mass size orbiting a supermassive black hole
is one of promising candidates for the source of gravitational waves.
Since the internal structure of such a compact object may be neglected
in this situation, we may adopt the black hole perturbation approach
with the compact object being regarded as a point particle.
In the black hole perturbation approach, we consider
the metric perturbation induced by a point particle of mass $\mu$
orbiting a black hole of mass $M$, where $\mu\ll M$.
At the lowest-order in the mass ratio ($\mu/M$), the motion of the
particle follows a geodesic of the background spacetime.
In the next order, however, the particle moves no longer
along a geodesic of the background because of its interaction with
the self-field.

Although this deviation from a background geodesic
is small for $\mu/M\ll1$ at each instant of time, after a large
lapse of time, it accumulates to become non-negligible.
For example, a circular orbit will not remain circular but
becomes a spiral-in orbit and the orbit
eventually plunges into the black hole.

If the time scale of the orbital evolution due to the self-force
is sufficiently long compared to the characteristic orbital time,
we may adopt the so-called adiabatic approximation in which
the orbit is assumed to be instantaneously geodesic with
the constants of motion changing very slowly with time.
In the Schwarzschild background case, we may assume the orbit to
lie on the equatorial plane and the geodesic motion
is determined by the energy and (the $z$-component of) angular momentum
of the particle.
In this case, the time variation of the energy and angular momentum
can be determined from the energy and angular momentum emitted
to infinity and absorbed into the black hole horizon
by using the conservation law.

However, there are cases when the adiabatic approximation breaks
down. For example, in the case of an extremely eccentric orbit
or an orbit close to the inner-most stable circular orbit,
the orbital evolution will not be adiabatic
because the stability of the orbit is strongly affected
by an infinitesimally small reaction force.
Furthermore, in the Kerr background, there is an additional
constant of motion, known as the Carter constant. Intuitively,
it describes the total orbital angular momentum, but unlike
the case of spherical symmetry, it has nothing to do with
the Killing vector field of the Kerr geometry.
The lack of its relation to the Killing vector makes us
impossible to evaluate the time change of the Carter constant
from the gravitational waves emitted to infinity and to event horizon,
even in the case when the adiabatic approximation is valid.
Thus it is in any case necessary to derive the self-force
of a particle explicitly.

The gravitational self-force $F^\mu$ is formally given as
\[
\frac{d^2 z^{\alpha}}{d\tau^2}
+\Gamma_{\mu\nu}^{\alpha}\frac{dz^{\mu}}{d\tau}\frac{dz^{\nu}}{d\tau}
=F^{\alpha} \,,
\]
where $\{z^{\alpha}(\tau)\}$ represents the orbit
with $\tau$ being the proper time measured in the background
geometry and $\Gamma_{\mu\nu}^{\alpha}$ is the
connection of the background.
The self-force arizes from the metric perturbation $h_{\mu\nu}$
induced by the particle:
\[
\tilde g_{\mu\nu} = g_{\mu\nu} + h_{\mu\nu}\,,
\]
and it is expressed as
\[
F^{\alpha}[h]=
-\mu P_{\beta}^{\alpha}
(\bar{h}_{\beta\gamma;\delta}
-\frac{1}{2}g_{\beta\gamma}
 {\bar{h}^{\epsilon}}{}_{\epsilon;\delta}
-\frac{1}{2}\bar{h}_{\gamma\delta;\beta}
+\frac{1}{4}g_{\gamma\delta}
{\bar{h}^{\epsilon}}{}_{\epsilon;\beta}
)u^{\gamma}u^{\delta}\,,
\]
where
${P_{\alpha}}^{\beta}={\delta_{\alpha}}^{\beta}+u_{\alpha}u^{\beta}$,
$\bar{h}_{\alpha\beta}=h_{\alpha\beta}
-\frac{1}{2}g_{\alpha\beta}h$
and $u^{\alpha}=dz^\alpha/d\tau$.

The metric perturbation diverges at the
location of the particle and so does the self-force.
Thus the above formal expression is in fact meaningless.
Fortunately, however, it is known that the metric perturbation
in the vicinity of the orbit can be divided into two parts
under the harmonic gauge condition; the direct part which has
support only on the past light-cone emanating from the field point
$x^\mu$ and the tail part which has support inside the past light-cone,
and the physical self-force is given by the tail part of the
metric perturbation which is regular as we let the field point
coincide with a point on the orbit; $x^\mu\to z^\mu(\tau)$
\cite{Mino:1996nk,Quinn:1999kj}.
It must be noted that the direct part can be evaluated by local analysis,
i.e., only with the knowledge of local geometrical quantities.
Therefore the physical self-force can be calculated as
\[
\lim_{x\to z(\tau)}F_{\alpha}[h^{\rm tail}(x)] =
\lim_{x\to z(\tau)}
\left(F_{\alpha}[h(x)]-F_{\alpha}[h^{{\rm dir}}(x)]\right).
\]
Furthermore, it has been revealed recently by Detweiler and
Whiting \cite{Detweiler:2002mi}
that the above devision of the metric can be slightly modified so that
the new direct part, called the S part, satisfies the same Einstein
equations as the full metric perturbation does, and the new tail part,
called the R part, satisfies the source-free Einstein equations,
and that the R part gives the
identical, regular self-force as the tail part does.
The important point is that the S part can be still
 evaluated locally near the
orbit without knowing the global solution.

When we perform this subtraction, we must evaluate the full self-force
and the direct part under the same gauge condition.
But the direct part is, by definition, defined only in the harmonic gauge.
On the other hand, the full metric perturbation is directly obtainable
only by the Regge-Wheeler-Zerilli or Teukolsky
formalism \cite{Regge:1957td,Zerilli:wd,Teukolsky:1973ha, Chrzanowski:wv}.
Therefore one must find a gauge transformation
that brings both the full metric perturbation and the direct part of the
metric perturbation to those in the same gauge.
This is called the {\it gauge problem} \footnote{Furthermore,
 the gravitational self-force is, because of the
equivalence principle, a gauge-variant notion. To give a genuinely
physical meaning to it, one must solve the second order
metric perturbation completely. This is however beyond the scope of
the present paper.}.

For the direct part, methods to obtain it under
the harmonic gauge condition were proposed
\cite{Mode-sum,Mino:2001mq,Barack:2001gx,Barack:2002mh}.
However, it seems extremely difficult to solve
the metric perturbation under the harmonic gauge
 because the metric components couple to each other in
a complicated way. This is one of the reasons
why the gauge problem is difficult to solve.

Recently, Barack and Ori \cite{Barack:2001ph} gave a useful
insight into the gauge problem. They proposed an intermediate gauge
approach in which only the direct part of the metric
in the harmonic gauge is
subtracted from the full metric perturbation in the RW gauge.
They then argued that the gauge-dependence of the self-force
is unimportant when averaged over a sufficiently long lapse of time.
Using this approach, the gravitational self-force for an orbit
plunging into a Schwarzschild black hole was calculated
by Barack and Lousto \cite{Barack:2002ku}.
But they also pointed out that the RW gauge is singular in the sense
that the resulting self-force will still have a direction-dependent
limit for general orbits.
The situation becomes worse in the Kerr background where
the only known gauge in which the metric perturbation can be
evaluated is the radiation gauge \cite{Chrzanowski:wv}, but the
metric perturbation becomes ill-defined in the neighborhood
of the particle, i.e., the Einstein equations are
not satisfied there \cite{Barack:2001ph}.

In this paper, as a direct approach to the gauge problem,
we consider a formalism to calculate the metric perturbation
in the harmonic gauge. We focus on the Schwarzschild background.
Instead of directly solving the metric perturbation in
the harmonic gauge, we consider the metric perturbation
in the RW gauge first, and then transform it
to the one in the harmonic gauge. Namely, we derive a set
of equations for gauge functions that transform the
metric perturbation in the RW gauge to the one in
the harmonic gauge.

The paper is organized as follows.
In Sec.~\ref{sec:form}, we formulate the gauge transformation
from the RW gauge to the harmonic gauge.
First we decompose the gauge transformation generators into the
Fourier-harmonics components.
Then the generators are devided into three parts;
the odd parity part and the even parity part which is further
devided into scalar and divergence free parts.
By the above procedure, we find a set of decoupled equations
for the gauge functions.
In Sec.~\ref{sec:discussion}, we summarize our formulation and
discuss remaining issues.
In Appendix \ref{app:RWZ}, we recapitulate the equations for the
Regge-Wheeler-Zerilli formalism by correcting typos in
Zerilli's paper \cite{Zerilli:wd}.

\section{Formulation}\label{sec:form}

We consider a metric perturbation $h_{\mu\nu}$
in the Schwarzschild background,
\begin{eqnarray}
g_{\mu\nu}dx^{\mu}dx^{\nu} =
-f(r)dt^2+f(r)^{-1}dr^2+r^2(d\theta^2+\sin^2\theta d\phi^2)\,;
\qquad f(r) = 1-\frac{2M}{r}\,.
\end{eqnarray}
We express the gauge transformation from the RW gauge
to the harmonic gauge as
\begin{eqnarray}
x_{{\rm RW}}^{\mu} &\rightarrow&
x_{{\rm H}}^{\mu} = x_{{\rm RW}}^{\mu} + \xi^{\mu}\,,
\label{displace} \\
h_{\mu\nu}^{{\rm RW}} &\rightarrow&
h_{\mu\nu}^{{\rm H}} =
h_{\mu\nu}^{{\rm RW}} - \xi_{\mu ;\nu} - \xi_{\nu ;\mu}\,,
\label{gauge-trans}
\end{eqnarray}
where the suffix RW stands for the RW gauge and
H for the harmonic gauge.

Substituting Eq.~(\ref{gauge-trans}) into the harmonic gauge
condition $\bar{h}_{\mu\nu}^{{\rm H}}{}^{;\nu}=0$,
we obtain the equations for $\xi^{\mu}$:
\begin{eqnarray}
\xi_{\mu}{}^{;\nu}{}_{;\nu} = \bar{h}^{\rm RW}_{\mu}{}^{\nu}{}_{;\nu}
=h^{\rm RW}_{\mu}{}^{\nu}{}_{;\nu}-\frac{1}{2}h^{\rm RW}_{;\mu}\,.
\label{eq:gauge-eq}
\end{eqnarray}
Using the static and spherical symmetry of the background,
we perform the Fourier-harmonic expansion of the above equation
and consider the equations for the expansion coefficients
for $\xi^{\mu}$ and $h^{\rm RW}_{\mu\nu}$.
We use the tensor harmonics introduced by Zerilli \cite{Zerilli:wd}
which are recapitulated in Appendix \ref{app:RWZ}.
Then according to the property under the parity transformation
$(\theta,\phi)\to(\pi-\theta,\phi+\pi)$,
we devide $\xi^\mu$ into the odd and even parity parts,
\begin{eqnarray}
\xi^\mu=\xi^\mu_{\rm (odd)}+\xi^\mu_{\rm (even)}\,,
\end{eqnarray}
and the even part is further decomposed into
the scalar and divergence-free part,
\begin{eqnarray}
\xi^{\mu}_{{\rm (even)}}=\xi^{;\mu} + \xi^{\mu}_{{\rm (v)}} \,,
\end{eqnarray}
where $\xi^\mu_{{\rm (v)}\,;\mu}=0$.

\subsection{Odd part}

First, we consider the odd part
which has the odd parity $(-1)^{\ell+1}$ under the parity transformation.
The gauge transformation generators and the metric perturbation are
given in the Fourier-harmonic expanded form as
\begin{eqnarray}
\xi_{\mu}^{{\rm (odd)}} &=&
\int d\omega \sum_{\ell m} e^{-i\omega t}
\Lambda_{\ell m\omega}(r)
\left\{ 0, 0, \frac{-1}{\sin\theta}\partial_{\phi}Y_{\ell m}(\theta,\phi),
\sin\theta\partial_{\theta}Y_{\ell m}(\theta,\phi) \right\},
\label{xi-decomp} \\
h_{\mu\nu}^{{\rm (odd)}} &=&
\int d\omega \sum_{\ell m}
\frac{\sqrt{2\ell(\ell+1)}}{r}e^{-i\omega t} \nonumber \\
& & \times
\left[ -h_{0\ell m\omega}(r){\bm c}_{\ell m\mu\nu}^{(0)}
+ih_{1\ell m\omega}(r){\bm c}_{\ell m\mu\nu}
+\frac{\sqrt{(\ell-1)(\ell+2)}}{2\,r}h_{2\ell m\omega}(r)
{\bm d}_{\ell m\mu\nu} \right],
\label{h-decomp}
\end{eqnarray}
where ${\bm c}_{\ell m\mu\nu}^{(0)}$,${\bm c}_{\ell m\mu\nu}$,
${\bm d}_{\ell m\mu\nu}$ are tensor harmonics
with odd parity \cite{Zerilli:wd}.
By substituting Eqs.~(\ref{xi-decomp}) and (\ref{h-decomp})
into Eq.~(\ref{eq:gauge-eq}), we obtain the equation for
$\Lambda_{\ell m\omega}(r)$:
\begin{eqnarray}
{\cal L}^{{\rm (odd)}}\Lambda_{\ell m\omega}(r) =
2\,R_{\ell m\omega}^{{\rm (odd)}}(r)
-\frac{16\,i\,\pi r^2}
{\sqrt{2\ell(\ell+1)(\ell-1)(\ell+2)}}D_{\ell m\omega}(r),
\label{lambda-eq}
\end{eqnarray}
where $R_{\ell m\omega}^{{\rm (odd)}}(r)$
is the Regge-Wheeler gauge invariant variable,
$D_{\ell m\omega}(r)$ is the Fourier-harmonic coefficient
of the stress-energy tensor, and the differential operator ${\cal L}$
is defined as
\begin{eqnarray}
{\cal L}^{{\rm (odd)}} &\equiv&
f(r)\frac{d^2}{dr^2}+f'(r)\frac{d}{dr}
+\left(\frac{\omega^2}{f(r)}-\frac{\ell(\ell+1)}{r^2}\right)
\nonumber \\
&=& {1 \over f(r)}\frac{d^2}{dr^{*\,2}}
+\left(\frac{\omega^2}{f(r)}-\frac{\ell(\ell+1)}{r^2}\right)
\,,
\label{eq:L}
\end{eqnarray}
where
$r^*=r+2M \log (r/2M-1)$ and $' = d/dr$.
The form of the differential operator ${\cal L}$ is slightly
different from the radial part of the scalar d'Alembertian,
but Eq.~(\ref{lambda-eq}) may be
solved by the standard Green function method.
The homogeneous solutions to construct the Green function
can be obtained by applying the method developed by Mano et al. \cite{Mano1}.
Here we note that the retarded causal boundary condition should be
imposed for the Green function.

\subsection{Even scalar part}

Next we consider the even scalar part of $\xi^{\mu}$ which
has the even parity $(-1)^{\ell}$ and is expressed by a gradient
of a scalar function $\xi$.
It is expressed as
\begin{eqnarray}
\xi_{\mu}^{{\rm (s)}} = \xi_{;\mu}\,; \quad
\xi=\int d\omega \sum_{\ell m}
\frac{1}{r}
\tilde{\xi}_{\ell m \omega}(r)
e^{-i\omega t}Y_{\ell m}(\theta,\phi) \,.
\end{eqnarray}

The gauge equation for $\xi_{\mu}^{{\rm (s)}}$ is derived from
$\xi_{\mu ;\nu}^{{\rm (s)}}{}^{;\nu}=J^{{\rm (s)}}_{;\mu}$,
which, because of the vanishing of the background Ricci tensor, gives
\begin{eqnarray}
\xi^{;\nu}{}_{;\nu}=J^{\rm (s)}\,,
\label{xieq}
\end{eqnarray}
where the source term $J^{\rm (s)}$ is determined from
the equation,
\begin{eqnarray}
{J^{{\rm (s)};\mu}}_{;\mu} &=&
\bar{h}^{\mu\nu}_{{\rm RW};\mu\nu} \,.
\label{Jseq}
\end{eqnarray}
The Fourier-harmonic expanded form of the above equation
is derived as follows.
First, we take the divergence of the metric perturbation
under the RW gauge;
\begin{eqnarray}
\bar{h}_{\mu\nu}^{{\rm RW};\nu} &=&
\int d\omega \sum_{\ell m} e^{-i\omega t}
\nonumber \\
&& \times
\left\{\left[-\frac{2f(r)}{r}H_{1\ell m\omega}^{{\rm RW}}(r)
-\frac{8\pi ir {B}_{\ell m\omega}^{{\rm (0)}}(r)}
{\sqrt{\ell(\ell+1)/2}}
-\frac{8\pi i\omega r^2
{F}_{\ell m\omega}(r)}
{\sqrt{\ell(\ell+1)(\ell-1)(\ell+2)/2}}
\right] Y_{\ell m}(\theta,\phi)({\bm e}_t)_{\mu}
\right.\nonumber \\
&& +\left[
\frac{2}{r}(H_{2\ell m\omega}^{{\rm RW}}(r)
-K_{\ell m\omega}^{{\rm RW}}(r))
+\frac{8\pi r {B}_{\ell m\omega}(r)}
{\sqrt{\ell(\ell+1)/2}}
-\frac{8\pi r^2
\frac{d}{dr} {F}_{\ell m\omega}(r)}
{\sqrt{\ell(\ell+1)(\ell-1)(\ell+2)/2}}
\right]
Y_{\ell m}(\theta,\phi)({\bm e}_r)_{\mu}
\nonumber \\
&& \left. +\frac{8\pi r^2}{\sqrt{\ell(\ell+1)(\ell-1)(\ell+2)/2}}
 {F}_{\ell m\omega}(r)({\bm e}_3)_{\mu} \right\} \,,
\label{hRWdiv}
\end{eqnarray}
where $({\bm e}_t)_{\mu}=(-1,0,0,0)$,
$({\bm e}_r)_{\mu}=(0,1,0,0)$ and
$({\bm e}_3)_{\mu} =
(0,0,\partial_{\theta}Y_{\ell m},\partial_{\phi}Y_{\ell m})$.
Then we take the divergence of the above once more to obtain
\begin{eqnarray}
\bar{h}_{\mu\nu}^{{\rm RW};\mu\nu} &=&
\int d\omega \sum_{\ell m} e^{-i\omega t}Y_{\ell m}(\theta,\phi)
\nonumber \\ && \times
\left[
-{\cal L}^{{\rm (s)}} \tilde{H}_{\ell m\omega}(r)
+8\pi\left( \frac{ {A}_{\ell m\omega}^{{\rm (0)}}(r)}{f(r)}
-f(r) {A}_{\ell m\omega}(r)
-\sqrt{2} {G}_{\ell m\omega}^{{\rm (s)}} \right)
\right]
\end{eqnarray}
where $\tilde{H}_{\ell m\omega}(r)$ is the trace of
the metric perturbation given by
\begin{eqnarray}
\tilde{H}_{\ell m\omega}(r) =
\frac{H^{\rm RW}_{0\ell m\omega}(r)
-H^{\rm RW}_{2\ell m\omega}(r)}{2}-K^{\rm RW}_{\ell m\omega}(r) \,.
\end{eqnarray}
The d'Alembertian of $J^{\rm(s)}$ is expanded as
\begin{eqnarray}
{J^{{\rm (s)};\mu}}_{;\mu} =
\int d\omega \sum_{\ell m} e^{-i\omega t}Y_{\ell m}(\theta,\phi)
{\cal L}^{{\rm (s)}}\tilde{J}^{{\rm (s)}}_{\ell m \omega}(r)\,,
\end{eqnarray}
where $\tilde J^{\rm(s)}_{\ell m\omega}$ is the Fourier-harmonic coefficient
of $J^{\rm(s)}$ and ${\cal L}^{\rm(s)}$ is the radial part of the scalar
d'Alembertian,
\begin{eqnarray}
{\cal L}^{{\rm (s)}} &\equiv&
f(r)\frac{d^2}{dr^2}+f'(r)\frac{d}{dr}
+\left(\frac{\omega^2}{f(r)}-\frac{\ell(\ell+1)}{r^2}
-\frac{f'(r)}{r}\right)
\nonumber\\
&=&{1 \over f(r)}\frac{d^2}{dr^{*\,2}}
+\left(\frac{\omega^2}{f(r)}-\frac{\ell(\ell+1)}{r^2}
-\frac{2\,M}{r^3}\right) \,.
\end{eqnarray}
Hence Eq.~(\ref{Jseq}) becomes
\begin{eqnarray}
{\cal L}^{{\rm (s)}}\tilde{J}^{{\rm (s)}}_{\ell m \omega}(r)
=
-{\cal L}^{{\rm (s)}} \tilde{H}_{\ell m\omega}(r)
+8\pi\left( \frac{ {A}_{\ell m\omega}^{{\rm (0)}}(r)}{f(r)}
-f(r) {A}_{\ell m\omega}(r)
-\sqrt{2} {G}_{\ell m\omega}^{{\rm (s)}}\right)
\,.
\label{tildeJseq}
\end{eqnarray}

Once the source term is obtained, we obtain from Eq.~(\ref{xieq})
the equation for $\tilde{\xi}_{\ell m \omega}(r)$ as
\begin{eqnarray}
{\cal L}^{{\rm (s)}}
\tilde{\xi}_{\ell m \omega}(r) &=&
r\tilde{J}^{{\rm (s)}}_{\ell m \omega}(r) \,.
\end{eqnarray}
Note that this equation as well as Eq.~(\ref{tildeJseq})
are just the radial part of the scalar
d'Alembertian, or equivalently the spin $s=0$ Teukolsky
equation, hence can be also solved by the Green function method.
The detail analysis of the homogeneous solutions
which satisfy the $s=0$ Teukolsky equation
is discussed in \cite{Mano1}.

\subsection{Even vector part}

Since the even vector part $\xi^{\mu}_{{\rm (v)}}$ satisfies
the divergence free condition, $\xi_{\mu}^{{\rm (v)}}{}^{;\mu}=0$,
this part has two degrees of freedom.
Therefore $\xi^\mu_{\rm(v)}$ is expressed in terms of two independent
radial functions,
\begin{eqnarray}
{\xi}_\mu^{{\rm (v)}}=\int d\omega
\sum_{\ell m}e^{-i\omega t}&
\biggl[&
\frac{1}{r}M_{0\ell m \omega}(r)Y_{\ell m}(\theta,\phi)({\bm e}_t)_\mu
+\frac{1}{r}M_{1\ell m \omega}(r)Y_{\ell m}(\theta,\phi)({\bm e}_r)_\mu
\nonumber \\ && \quad
+\frac{1}{\ell(\ell+1)}\left\{\frac{-i\omega r}{f(r)} M_{0\ell m \omega}(r)
+\partial_r \left( rf(r)M_{1\ell m \omega}(r) \right)\right\}
({\bm e}_3)_\mu
\biggr]\,.
\end{eqnarray}

To solve the gauge transformation equation (\ref{eq:gauge-eq})
for this part,
we introduce an auxiliary field
$F_{\mu\nu}:=\xi_{\nu;\mu}^{{\rm (v)}}-\xi_{\mu;\nu}^{{\rm (v)}}$.
Then we have
\begin{eqnarray}
F_{\mu \nu}{}^{;\nu}
=- J^{{\rm (v)}}_{\mu} \,;
\quad
J^{{\rm (v)}}_{\mu}\equiv
\bar{h}_{\mu\nu}^{{\rm RW};\nu}-J^{{\rm (s)}}_{;\mu}
\,,
\end{eqnarray}
where we have again used the fact that the background
Ricci tensor vanishes. Note that the Fourier-harmonic expansion
of $\bar{h}_{\mu\nu}^{{\rm RW};\nu}$ is given by Eq.~(\ref{hRWdiv}).
The above equation is the Maxwell equation,
so it can be solved by using the
Teukolsky formalism \cite{Teukolsky:1973ha} for the electromagnetic field ($s=\pm 1$).
Namely, we introduce the following variables:
\begin{eqnarray}
\phi_0=F_{\mu\nu}l^{\mu}m^{\nu} \,, \quad
\phi_2=F_{\mu\nu}m^{*\mu}n^{\nu} \,,
\label{eq:Fmunu}
\end{eqnarray}
where $*$ denotes the complex conjugate, and
$l^{\mu}$, $n^{\mu}$, $m^{\mu}$ are the Kinnersley null tetrad
defined by
\begin{eqnarray}
l^{\mu}=\left\{\frac{1}{f(r)},1,0,0\right\} \,, \quad
n^{\mu}=\left\{\frac{1}{2},-\frac{f(r)}{2},0,0\right\}\,, \quad
m^{\mu}=\frac{1}{\sqrt{2}r}\left\{0,0,1,\frac{i}{\sin\theta}\right\}\,.
\end{eqnarray}
The variables $\phi_0$ and $\phi_2$
satisfy the $s=\pm1$ Teukolsky equation,
\begin{eqnarray}
{\cal L}^{{\rm (Teuk)}}_s \, \Psi_s = -4\pi r^2 T_s \,,
\label{eq:Teuk}
\end{eqnarray}
where $\Psi_1=\phi_0$ and $\Psi_{-1}=\phi_2/r^2$, and
${\cal L}^{{\rm (Teuk)}}_s$ is the Teukolsky differential
operator defined as
\begin{eqnarray}
{\cal L}^{{\rm (Teuk)}}_s &:=&
-\frac{r^2}{f(r)}\partial_t^2+2s\left(\frac{M}{f(r)}-r\right)\partial_t
+\left(r^2f(r)\right)^{-s}
\partial_r\left[\left(r^2f(r)\right)^{s+1}\partial_r\right] \nonumber \\
& &
+\frac{1}{\sin\theta}\partial_{\theta}(\sin\theta\partial_{\theta})
+\frac{1}{\sin^2\theta}\partial^2_{\phi}
+\frac{2is\cos\theta}{\sin^2\theta}\partial_{\phi}
-s(s\cot^2\theta-1) \,.
\end{eqnarray}
The source terms $T_s$, which are calculated from
$J^{{\rm (v)}}_{\mu}$, are
\begin{eqnarray}
T_{1} &=& {1\over \sqrt{2}r}
\left(\partial_{\theta}+{i\over \sin \theta}\partial_{\phi} \right)
\left[J^{{\rm (v)}}_{\mu}\,l^{\mu}\right]
-\left(\partial_r-{i\omega \over f(r)} +{3 \over r}\right)
\left[J^{{\rm (v)}}_{\mu}\,m^{\mu}\right]
\,, \\
T_{-1} &=& -{1\over \sqrt{2}r}
\left(\partial_{\theta}-{i\over \sin \theta}\partial_{\phi} \right)
\left[J^{{\rm (v)}}_{\mu}\,n^{\mu}\right]
-{f(r) \over 2}\left(\partial_r+{i\omega \over f(r)} +{3 \over r}\right)
\left[J^{{\rm (v)}}_{\mu}\,m^{*\mu}\right]
\,.
\end{eqnarray}
The solution of this Teukolsky equation is also
analyzed in \cite{Mano1}.

Once we obtain $\phi_0$ and $\phi_2$, the two radial functions for
the gauge transformation are calculated from
Eq.~(\ref{eq:Fmunu}) which reads
\begin{eqnarray}
\phi_0 &=&\int d\omega
\sum_{\ell m} e^{-i\omega t}
\frac{-\,{}_{1}Y_{\ell m}(\theta,\phi)}{\sqrt{2\ell(\ell +1)}rf(r)}
\biggl[\frac{\ell(\ell+1)}{r}(M_{0\ell m \omega}(r)-f(r)M_{1\ell m \omega}(r))
\nonumber \\
& & -i\omega\left\{\frac{-i\omega r}{f(r)}
M_{0\ell m \omega}(r)+\frac{d}{dr} (rf(r)M_{1\ell m \omega}(r))\right\}
+f(r)\frac{d}{dr}\left\{\frac{-i\omega r}{f(r)}
M_{0\ell m \omega}(r)+\frac{d}{dr}
(rf(r)M_{1\ell m \omega}(r))\right\}
\biggr] \,, \\
\phi_2 &=&\int d\omega
\sum_{\ell m} e^{-i\omega t}
\frac{{}_{-1}Y_{\ell m}(\theta,\phi)}{2\sqrt{2\ell(\ell +1)}r}
\biggl[-\frac{\ell(\ell+1)}{r}(M_{0\ell m \omega}(r)
+f(r)M_{1\ell m \omega}(r))
\nonumber \\
& & +i\omega\left\{\frac{-i\omega r}{f(r)}M_{0\ell m \omega}(r)
+\frac{d}{dr}(rf(r)M_{1\ell m \omega}(r))\right\}
+f(r)\frac{d}{dr}\left\{\frac{-i\omega r}{f(r)}M_{0\ell m \omega}(r)
+\frac{d}{dr}(rf(r)M_{1\ell m \omega}(r))\right\}
\biggr] \,,
\end{eqnarray}
where ${}_{s}Y_{\ell m}(\theta,\phi)$ are the
spin-weighted spherical harmonics.
Performing the Fourier and spin-weighted spherical harmonic expansion
for $\phi_0$ and $\phi_2$,
\begin{eqnarray}
\phi_0&=&\int d\omega
\sum_{\ell m}
\tilde{\phi}_{0\ell m \omega}(r) e^{-i\omega t}
{}_{1}Y_{\ell m}(\theta,\phi) \,,
\\
\phi_2&=&\int d\omega
\sum_{\ell m}
\tilde{\phi}_{2\ell m \omega}(r) e^{-i\omega t}
{}_{-1}Y_{\ell m}(\theta,\phi) \,,
\end{eqnarray}
we obtain the equations,
\begin{eqnarray}
\tilde{\phi}_{0\ell m \omega}(r)
+\frac{2}{f(r)}\tilde{\phi}_{2\ell m \omega}(r) &=&
-\frac{\sqrt{2/\ell(\ell +1)}}{rf(r)}
\left[\left(\frac{\ell(\ell +1)}{r}-\frac{\omega^2 r}{f(r)}
\right)M_{0\ell m \omega}(r)
-i\omega\frac{d}{dr}(rf(r)M_{1\ell m \omega}(r))
\right]\,, \label{eq:Meq-1}\\
\tilde{\phi}_{0\ell m \omega}(r)
-\frac{2}{f(r)}\tilde{\phi}_{2\ell m \omega}(r) &=&
\frac{\sqrt{2/\ell(\ell +1)}}{rf(r)}
\left[\frac{\ell(\ell +1)f(r)}{r}M_{1\ell m \omega}(r)
\right. \nonumber \\
& & \left.
-f(r)\frac{d^2}{dr^2}(rf(r)M_{1\ell m \omega}(r))
+i\omega f(r)\frac{d}{dr}
\left( \frac{r}{f(r)}M_{0\ell m \omega}(r)\right)
\right]. \label{eq:Meq-2}
\end{eqnarray}

We can eliminate $M_{1\ell m \omega}(r)$ from Eqs.~(\ref{eq:Meq-1})
and (\ref{eq:Meq-2}) to obtain a decoupled
equation for $M_{0\ell m \omega}(r)$,
\begin{eqnarray}
{\cal L}^{{\rm (s)}} M_{0\ell m \omega}(r) &=&
-\frac{r^2}{\sqrt{2\ell(\ell+1)}}
\left[f(r)^2\frac{d^2}{dr^2}\tilde{\phi}_{0\ell m\omega}(r)
+f(r)\left(\frac{2+2f(r)}{r}+f'(r)+i\omega\right)
\frac{d}{dr}\tilde{\phi}_{0\ell m\omega}(r)
\right. \nonumber \\
& &
+\frac{rf'(r)-(\ell-1)(\ell+2)f(r)+i\omega r(1+2f(r))}{r^2}
\tilde{\phi}_{0\ell m\omega}(r)
+2f(r)\frac{d^2}{dr^2}\tilde{\phi}_{2\ell m\omega}(r)
\nonumber \\
& & \left.
+2\left(\frac{2+2f(r)}{r}-f'(r)-i\omega\right)
\frac{d}{dr}\tilde{\phi}_{2\ell m\omega}(r)
-2\frac{rf'(r)+(\ell-1)(\ell+2)+3i\omega r}{r^2}
\tilde{\phi}_{2\ell m\omega}(r)
\right] , \\
&=&
\frac{2 r^2 f(r)}{\sqrt{2\ell(\ell+1)}}\left[
4\pi\left(2r^2 \tilde{T}_{-1\ell m\omega}(r)
+f(r)\tilde{T}_{1\ell m\omega}(r)\right)
\right .\nonumber \\
& &
-f(r)(i\omega+f'(r))\frac{d}{dr}\tilde{\phi}_{0\ell m\omega}(r)
-\frac{(1+i\omega r)(rf'(r)+i\omega r)}{r^2}
\tilde{\phi}_{0\ell m\omega}(r)
\nonumber \\
& & \left.
+\frac{2}{r}(ir\omega-1-7f(r))
\frac{d}{dr}\tilde{\phi}_{2\ell m\omega}(r)
+2\left(\frac{\omega^2+i\omega/r}{f(r)}+\frac{5f(r)-1}{r^2}\right)
\tilde{\phi}_{2\ell m\omega}(r)\right],
\label{eq:M1}
\end{eqnarray}
where $\tilde{T}_{s\ell m\omega}(r)$
is the Fourier-harmonic coefficient of the source term in
the Teukolsky equation (\ref{eq:Teuk}).
Interestingly, Eq.~(\ref{eq:M1}) has the same form as
the $s=0$ Teukolsky equation.
Once we obtain $M_{0\ell m \omega}(r)$ by solving the above equation,
$M_{1\ell m \omega}(r)$ is derived from
\begin{eqnarray}
M_{1\ell m \omega}(r)&=&
\frac{-ir^2}{\omega\sqrt{2\ell(\ell+1)}}
\left[ f(r)\frac{d}{dr}\tilde{\phi}_{0\ell m\omega}(r)
+\frac{1+i\omega r}{r}\tilde{\phi}_{0\ell m\omega}(r)
+2\frac{d}{dr}\tilde{\phi}_{2\ell m\omega}(r)
+2\frac{f(r)-i\omega r}{r f(r)}\tilde{\phi}_{2\ell m\omega}(r)
\right] \nonumber \\
& &-\frac{i}{\omega}\left( \frac{d}{dr}M_{0\ell m\omega}(r)
-\frac{M_{0\ell m\omega}(r)}{r}\right),
\end{eqnarray}
which also follows from Eqs.~(\ref{eq:Meq-1}) and (\ref{eq:Meq-2}).

\section{Summary and Discussion}\label{sec:discussion}

In this paper, to solve the gauge problem of the gravitational
self-force, we have considered the gauge transformation
from the Regge-Wheeler gauge to the harmonic gauge
and have presented a formalism to obtain the infinitesimal
displacement vector of this transformation, $\xi^{\mu}$.
First, we have performed the Fourier-harmonic expansion
of $\xi^{\mu}$ and divided it into the odd and even parity parts.
The odd part has only one degree of freedom and
it turns out that the gauge transformation can be found
by solving a single second-order differential equation
for the radial function.
As for the even parity part, we have further divided it
into scalar and vector parts where the scalar part is given
by the gradient of a scalar function and the vector part
is divergence-free. The scalar part has by definition
only one degree of freedom, and we have found
that it can be obtained by solving two second-order differential
equations consecutively. These two equations are found to be
identical to the $s=0$ Teukolsky equation.
The vector part has two degrees of freedom, and the gauge
transformation equations give equations that are coupled
in a complicated way. However, by introducing
two auxiliary variables which satisfy the $s=\pm1$
Teukolsky equations, we have succeeded in deriving
a decoupled second-order equation for one of the
gauge functions with the source term given by the auxiliary
variables. Interestingly, this second-order equation has the
same form as the $s=0$ Teukolsky equation.
The other gauge function is then simply
given by applying a differential operator to the first.

Since all the equations to be solved have the
form analogous to or equal to the Regge-Wheeler equation,
we can derive analytic expressions for
their homogeneous solutions by using the
 Mano-Suzuki-Takasugi method \cite{Mano1}, and
construct the Green function from these homogeneous solutions.
So we conclude that the gauge transformation can be
solved by using the Green function method,
and we can construct the metric perturbation in the harmonic gauge.
In practice, however, it may not be easy to solve for the
gauge transformation since
it involves products of Green functions with double integrals.
Derivation of the gauge transformation functions in a closed,
practically tractable form is left for future study.

Another approach to the gauge problem is to
consider the self-force in a gauge different from the
harmonic gauge, similar to (but very different in principle from)
the intermediate gauge approach
proposed by Barack and Ori \cite{Barack:2001ph}.
Here the recent result by Detweiler and Whiting \cite{Detweiler:2002mi}
becomes crucial. Their observation that the S part
and the R part play the identical roles as the direct part
and the tail part, respectively, and that the S part satisfies
the same inhomogeneous Einstein equations as the full metric
perturbation enables us to define the S part and the R part
of the metric perturbation unambiguously in an arbitrary gauge
as long as the gauge condition is consistent with
the Einstein equations. For example, given the S part of the
metric perturbation in the harmonic gauge, one can perform
the gauge transformation of it to the RW gauge and
the resulting metric perturbation which satisfies the Einstein
equations can be identified as the S part of the metric
perturbation in the RW gauge. Then, after solving
the Regge-Wheeler-Zerilli equations to obtain the full metric
perturbation, it is straightforward to derive the R part of
the metric perturbation in the RW gauge \cite{Mino}.
The calculation of the self-force in the RW gauge
in this manner is in progress \cite{SNS2}.

Finally, we comment on the self-force in the case of the Kerr background.
In the Schwarzschild case, it was possible to use the
Regge-Wheeler-Zerilli formalism to obtain the metric perturbation
in the RW gauge. However, in the Kerr case, there is no known gauge
in which the full metric perturbation can be calculated.
The Chrzanowski method \cite{Chrzanowski:wv} based on the Teukolsky formalism
can give the metric perturbation in the (ingoing or outgoing) radiation
gauge, but only outside the range of radial coordinates the
orbit resides in.
One possible way to circumvent this difficulty is to consider
first the regularization of the Weyl scalar $\Psi_4$.
Given an orbit, $\Psi_4$ can be calculated
by the Teukolsky formalism, and the S part of it, $\Psi_4^{\rm S}$,
can be calculated from the S part of
the metric perturbation in the harmonic gauge,
$h_{\mu\nu}^{\rm S,H}$,
\begin{eqnarray}
\Psi_4^{\rm S} = \hat{\Psi}_4[h_{\mu\nu}^{\rm S,H}] \,,
\end{eqnarray}
where $\hat{\Psi}_4$ is the operator to derive the Wely scalar
from a given metric perturbation.
Then the R part of $\Psi_4$ can be derived
by subtracting the S part from the Weyl scalar,
\begin{eqnarray}
\Psi_4^{\rm R} = \Psi_4 -\Psi_4^{\rm S} \,.
\end{eqnarray}
Now $\Psi_4^{\rm R}$ satisfies the homogeneous Teukolsky equation.
Hence using the Chrzanowski method,
we may construct the R part of the metric perturbation in the
radiation gauge and derive the self-force.
Since this procedure involves many derivative operations,
the metric perturbation $h_{\mu\nu}^{\rm S,H}$ has to be evaluated to
with a sufficiently high accuracy which may be practically a
difficult task, if not impossible. Feasibility of this method
should surely be investigated.

\acknowledgements

We would like to thank T.~Tanaka and Y.~Mino
for useful discussions.
HN and MS would like to thank all the participants at
the Radiation Reaction Focus Session and
the 5th Capra Ranch Meeting, PSU, State College in U.S.A.,
for invaluable discussions.
This work was supported in part by Monbukagaku-sho Grant-in-Aid
for Scientific Research Nos.~1047214 and 12640269, and by
the Center for Gravitational Wave
Physics, which is funded by the National Science Foundation under
Cooperative Agreement PHY 0114375.

\begin{appendix}

\section{The field equation on the Regge-Wheeler gauge} \label{app:RWZ}

In this appendix, we recapitulate the equations for
the Regge-Wheeler-Zerilli formalism. In doing so, we correct
some minor errors in Zerilli's paper \cite{Zerilli:wd}.
Here an equation number given as (Z:1) denotes the equation (1)
in Zerilli's paper for comparison,
and a label $[CRTD]$ to an equation means
it is corrected.

We consider the linearized Einstein equations
for the perturbed metric,
\[
\tilde g_{\mu\nu}=g_{\mu\nu}+h_{\mu\nu}\,,
\]
where $g_{\mu\nu}$ is the background metric.
Then the Einstein tensor and the stress-energy tensor up to
the linear order can be expanded as
\begin{eqnarray}
G_{\mu\nu}[\tilde g_{\mu\nu}] &=&
{G}_{\mu\nu}[ g_{\mu\nu}]
+\delta G_{\mu\nu}[h_{\mu\nu}]
+O(h^2), \\
\tilde T_{\mu\nu} &=&
T_{\mu\nu}+\delta T_{\mu\nu},
\end{eqnarray}
where $f_{\mu}={h_{\mu\alpha}}^{;\alpha}$ and
\begin{eqnarray}
\delta G_{\mu\nu}[h_{\mu\nu}] &=&
-\frac{1}{2}h_{\mu\nu;\alpha}{}^{;\alpha}+f_{(\mu;\nu)}
-R_{\alpha\mu\beta\nu}h^{\alpha\beta}
-\frac{1}{2}h_{;\mu;\nu}
+R^{\alpha}{}_{(\mu}h_{\nu)\alpha}
\nonumber\\
 & &
-\frac{1}{2} g_{\mu\nu}
(f_{\lambda}{}^{;\lambda}-h_{;\lambda}{}^{;\lambda})
-\frac{1}{2}h_{\mu\nu}R
+\frac{1}{2}g_{\mu\nu}h_{\alpha\beta}R^{\alpha\beta} \,.
\end{eqnarray}
When the background is Ricci flat, $R_{\mu\nu}^{{\rm (b)}}=0$,
the above equation is rewritten as
\begin{eqnarray}
-\frac{1}{2}h_{\mu\nu;\alpha}{}^{;\alpha}+f_{(\mu;\nu)}
-R_{\alpha\mu\beta\nu}h^{\alpha\beta}
-\frac{1}{2}h_{;\mu;\nu}
-\frac{1}{2}g_{\mu\nu}
(f_{\lambda}{}^{;\lambda}-h_{;\lambda}{}^{;\lambda}) =
8\pi\delta T_{\mu\nu}.
\label{eq:lineareq}
\end{eqnarray}

We apply the above to the case of the Schwarzschild background, and
expand $h_{\mu\nu}$ ((Z:D2a) and (Z:D2b))
and $\delta T_{\mu\nu}$ in tensor harmonics,
\begin{eqnarray}
\bm{h} &=& \sum_{\ell m} \left[
f(r)H_{0\ell m}(t,r)\bm{a}^{(0)}_{\ell m}
-i\sqrt{2}H_{1\ell m}(t,r)\bm{a}^{(1)}_{\ell m}
+\frac{1}{f(r)}H_{2\ell m}(t,r)\bm{a}_{\ell m}
\right. \nonumber \\
& &
-\frac{i}{r}\sqrt{2\ell(\ell+1)}h^{(e)}_{0\ell m}(t,r)\bm{b}^{(0)}_{\ell m}
+\frac{1}{r}\sqrt{2\ell(\ell+1)}h^{(e)}_{1\ell m}(t,r)\bm{b}_{\ell m}
\nonumber \\
& &
+\sqrt{\frac{1}{2}\ell(\ell+1)(\ell-1)(\ell+2)}G_{\ell m}(t,r)\bm{f}_{\ell m}
+\left(\sqrt{2}K_{\ell m}(t,r)
 -\frac{\ell(\ell+1)}{{\sqrt{2}}}G_{\ell m}(t,r)\right)\bm{g}_{\ell m}
\nonumber \\
& &
\left.
-\frac{\sqrt{2\ell(\ell+1)}}{r}h_{0\ell m}(t,r)\bm{c}^{(0)}_{\ell m}
+\frac{i\sqrt{2\ell(\ell+1)}}{r}h_{1\ell m}(t,r)\bm{c}_{\ell m}
\right.
\nonumber \\
& &
\left.
+\frac{\sqrt{2\ell(\ell+1)(\ell-1)(\ell+2)}}{2r}h_{2\ell m}(t,r)\bm{d}_{\ell m}
\right] \quad [CRTD] \,, \label{eq:hharm}
\\
\delta \bm{T} &=& \sum_{\ell m}\left[
A_{\ell m}^{(0)}\bm{a}_{\ell m}^{(0)}
+A_{\ell m}^{(1)}\bm{a}_{\ell m}^{(1)}
+A_{\ell m}\bm{a}_{\ell m}
+B_{\ell m}^{(0)}\bm{b}_{\ell m}^{(0)}
+ B_{\ell m}\bm{b}_{\ell m}
+Q_{\ell m}^{(0)}\bm{c}_{\ell m}^{(0)}
+Q_{\ell m}\bm{c}_{\ell m}
 \right. \nonumber \\
& & \hspace{7cm} \left.
+D_{\ell m}\bm{d}_{\ell m}
+G_{\ell m}^{(s)}\bm{g}_{\ell m}
+F_{\ell m}\bm{f}_{\ell m}
\right]\,,
\label{eq:Tharm}
\end{eqnarray}
where
we use $h^{(e)}_{0\ell m}$ and
$h^{(e)}_{1\ell m}$ for the even part coefficients instead of
$h^{(m)}_{0\ell m}$ and $h^{(m)}_{0\ell m}$, respectively,
in Zerilli's paper, and the coefficient $G_{\ell m}^{(s)}$
instead of the Zerilli's notation $G_{\ell m}$
for the energy-momentum tensor, and
$\bm{a}_{\ell m}^{(0)}$, $\bm{a}_{\ell m}\,,\cdots$ are the
ten tensor harmonics (Z:A2a-j) defined as
\begin{eqnarray}
{\bf a}_{\ell m}{}^{(0)}&=&
\left(\begin{array}{cccc}
Y_{\ell m} & 0 & 0 & 0 \\
0 & 0 & 0 & 0 \\
0 & 0 & 0 & 0 \\
0 & 0 & 0 & 0
\end{array}\right) \,,
\\
{\bf a}_{\ell m}{}^{(1)}&=&(i/\sqrt{2})
\left(\begin{array}{cccc}
0 & Y_{\ell m} & 0 & 0 \\
Sym & 0 & 0 & 0 \\
0 & 0 & 0 & 0 \\
0 & 0 & 0 & 0
\end{array}\right)\,,
\\
{\bf a}_{\ell m}&=&
\left(\begin{array}{cccc}
0 & 0 & 0 & 0 \\
0 & Y_{\ell m} & 0 & 0 \\
0 & 0 & 0 & 0 \\
0 & 0 & 0 & 0
\end{array}\right)\,,
\\
{\bf b}_{\ell m}{}^{(0)}&=&
ir[2\ell(\ell+1)]^{-1/2}
\left(\begin{array}{cccc}
0 & 0 & (\partial /\partial \theta)Y_{\ell m}
                             & (\partial /\partial \phi)Y_{\ell m} \\
0 & 0 & 0 & 0 \\
Sym & 0 & 0 & 0 \\
Sym & 0 & 0 & 0
\end{array}\right)\,,
\\
{\bf b}_{\ell m}&=&
r[2\ell(\ell+1)]^{-1/2}
\left(\begin{array}{cccc}
0 & 0 & 0 & 0 \\
0 & 0 & (\partial/\partial \theta)Y_{\ell m}
                              &(\partial /\partial \phi)Y_{\ell m} \\
0 & Sym & 0 & 0 \\
0 & Sym & 0 & 0
\end{array}\right)\,,
\\
{\bf c}_{\ell m}{}^{(0)}&=&r[2\ell(\ell+1)]^{-1/2}
\left(\begin{array}{cccc}
0 & 0 & (1/\sin \theta)(\partial /\partial \phi)Y_{\ell m}
& -\sin \theta(\partial /\partial \theta)Y_{\ell m} \\
0 & 0 & 0 & 0 \\
Sym & 0 & 0 & 0 \\
Sym & 0 & 0 & 0
\end{array}\right)\,,
\\
{\bf c}_{\ell m}&=&ir[2\ell(\ell+1)]^{-1/2}
\left(\begin{array}{cccc}
0 & 0 & 0 & 0 \\
0 & 0 & (1/\sin \theta)(\partial /\partial \phi)Y_{\ell m}
& -\sin \theta(\partial /\partial \theta)Y_{\ell m} \\
0 & Sym & 0 & 0 \\
0 & Sym & 0 & 0
\end{array}\right)\,,
\\
{\bf d}_{\ell m}&=&ir^2[2\ell(\ell+1)(\ell-1)(\ell+2)]^{-1/2}
\left(\begin{array}{cccc}
0 & 0 & 0 & 0 \\
0 & 0 & 0 & 0 \\
0 & 0 & -(1/\sin \theta)X_{\ell m} & \sin \theta W_{\ell m} \\
0 & 0 & Sym & \sin \theta X_{\ell m}
\end{array}\right)\,,
\\
{\bf g}_{\ell m}&=&(r^2/\sqrt{2})
\left(\begin{array}{cccc}
0 & 0 & 0 & 0 \\
0 & 0 & 0 & 0 \\
0 & 0 & Y_{\ell m} & 0 \\
0 & 0 & 0 & \sin ^2\theta Y_{\ell m}
\end{array}\right)\,,
\\
{\bf f}_{\ell m}&=&r^2[2\ell(\ell+1)(\ell-1)(\ell+2)]^{-1/2}
\left(\begin{array}{cccc}
0 & 0 & 0 & 0 \\
0 & 0 & 0 & 0 \\
0 & 0 & W_{\ell m} & X_{\ell m} \\
0 & 0 & Sym & -\sin ^2\theta W_{\ell m}
\end{array}\right) \,[CRTD] \,.
\end{eqnarray}
Here the angular functions
$X_{\ell m}$ and $W_{\ell m}$ are given by
\begin{eqnarray}
X_{\ell m}&=&2{\partial \over \partial \phi}
\left({\partial \over \partial \theta}-\cot \theta \right)Y_{\ell m} \,, \\
W_{\ell m}&=&\left({\partial ^2\over \partial \theta ^2}
-\cot \theta {\partial \over \partial \theta}
-{1 \over \sin ^2 \theta}
{\partial ^2 \over \partial \phi ^2}Y_{\ell m}
\right) \,.
\end{eqnarray}

For a point particle moving along a geodesic,
the stress-energy tensor takes the form,
\begin{eqnarray}
T^{\mu\nu}&=& \mu \int^{+\infty}_{-\infty} \delta^{(4)}(x-z(\tau))
{dz^{\mu} \over d\tau}{dz^{\nu} \over d\tau}d\tau \nonumber \\
&=& \mu \, \gamma {dz^{\mu} \over dt}{dz^{\nu} \over dt}
{\delta(r-R(t)) \over r^2} \delta ^{(2)}(\Omega-\Omega (t)) \,,
\end{eqnarray}
where the following notation for the particle orbit is used.
\begin{eqnarray}
z^{\mu}&=& z^{\mu}(\tau)=\{T(\tau),R(\tau),\Theta(\tau),\Phi(\tau)\} \,,
\\
\gamma &=& {dT(\tau) \over d\tau} \,.
\end{eqnarray}
This stress-energy tensor is expressed in terms of the tensor harmonics
as given in Table I (correspond to (Z:Table III)).

\begin{table}
\begin{center}
 \caption{Stress-Energy Tensor in terms of Tensor Harmonics}
\renewcommand{\arraystretch}{1.8}
  \begin{tabular}{|c|c|c|}
\hline
Description & Dependence of "driving term" on $r$ and $t$
& Tensor harmonic \\
    \hline
Even
& $\displaystyle
A_{\ell m}(r,t)=\mu \gamma \left({dR \over dt}\right)^2(r-2M)^{-2}
\delta(r-R(t))Y_{\ell m}^*(\Omega (t))$
& ${\bm a}_{\ell m}(\theta,\phi)$ \\ \hline
Even
& $\displaystyle
A_{\ell m}^{(0)}=\mu \gamma \left(1-{2M \over r}\right)^2r^{-2}
\delta(r-R(t))Y_{\ell m}^*(\Omega (t))$
& ${\bm a}_{\ell m}^{(0)}(\theta,\phi)$ \\ \hline
Even
& $\displaystyle
A_{\ell m}^{(1)}=-\sqrt{2}i\mu \gamma {dR \over dt}r^{-2}
\delta(r-R(t))Y_{\ell m}^*(\Omega (t))\quad[CRTD]$
& ${\bm a}_{\ell m}^{(1)}(\theta,\phi)$ \\ \hline
Even
& $\displaystyle
B_{\ell m}^{(0)}=-[{1 \over 2}\ell(\ell+1)]^{-1/2}i\mu \gamma
\left(1-{2M \over r}\right)r^{-1}
\delta(r-R(t))dY_{\ell m}^*(\Omega (t))/dt\quad[CRTD]$
& ${\bm b}_{\ell m}^{(0)}(\theta,\phi)$ \\ \hline
Even
& $\displaystyle
B_{\ell m}=[{1 \over 2}\ell(\ell+1)]^{-1/2}\mu \gamma
(r-2M)^{-1}{dR \over dt}
\delta(r-R(t))dY_{\ell m}^*(\Omega (t))/dt$
& ${\bm b}_{\ell m}(\theta,\phi)$ \\ \hline
Odd
& $\displaystyle
Q_{\ell m}^{(0)}=[{1 \over 2}\ell(\ell+1)]^{-1/2}\mu \gamma
\left(1-{2M \over r}\right)r^{-1}
\delta(r-R(t))
$
& ${\bm c}_{\ell m}^{(0)}(\theta,\phi)$ \\
 &
$\displaystyle
\times \left[{1\over \sin \Theta}{\partial Y_{\ell m}^* \over \partial \Phi}
{d \Theta \over dt}-\sin \Theta{\partial Y_{\ell m}^* \over \partial \Theta}
{d\Phi \over dt}\right]$
& $$ \\ \hline
Odd
& $\displaystyle
Q_{\ell m}=-[{1 \over 2}\ell(\ell+1)]^{-1/2}i\mu \gamma {dR \over dt}
(r-2M)^{-1}
\delta(r-R(t))
$
& ${\bm c}_{\ell m}(\theta,\phi)$ \\
 &
$\displaystyle
\times \left[{1\over \sin\Theta}{\partial Y_{\ell m}^* \over \partial \Phi}
{d \Theta \over dt}-\sin \Theta{\partial Y_{\ell m}^* \over \partial \Theta}
{d\Phi \over dt}\right]\quad[CRTD]$
& $$ \\ \hline
Odd
& $\displaystyle
D_{\ell m}=-[{1 \over 2}\ell(\ell+1)(\ell-1)(\ell+2)]^{-1/2}i\mu \gamma
\delta(r-R(t))$
& ${\bm d}_{\ell m}(\theta,\phi)$ \\
 &
$\displaystyle
\times \left({1 \over 2}\left[({d \Theta \over dt})^2
-\sin^2\Theta ({d\Phi \over dt})^2\right]
{1 \over \sin \Theta}X_{\ell m}^*[\Omega(t)]
-\sin \Theta {d\Phi \over dt}{d \Theta \over dt}
  W_{\ell m}^*[\Omega(t)]\right)$
& $$ \\ \hline
Even
& $\displaystyle
F_{\ell m}=[{1 \over 2}\ell(\ell+1)(\ell-1)(\ell+2)]^{-1/2}\mu \gamma
\delta(r-R(t))$
& ${\bm f}_{\ell m}(\theta,\phi)$ \\
 &
$\displaystyle
\times \left({d\Phi \over dt}{d \Theta \over dt}X_{\ell m}^*[\Omega(t)]
+{1 \over 2}\left[({d \Theta \over dt})^2
-\sin^2\Theta ({d\Phi \over dt})^2\right]W_{\ell m}^*[\Omega(t)]\right)$
& $$ \\ \hline
Even
& $\displaystyle
G_{\ell m}^{(s)}={\mu \gamma \over \sqrt{2}}\delta(r-R(t))
\left[({d \Theta \over dt})^2
+\sin^2\Theta ({d\Phi \over dt})^2\right]Y_{\ell m}^*(\Omega (t))$
& ${\bm g}_{\ell m}(\theta,\phi)$ \\
\hline
  \end{tabular}
 \end{center}
\end{table}

Substituting Eqs.~(\ref{eq:hharm}) and (\ref{eq:Tharm}) into
Eq.~(\ref{eq:lineareq}),
we obtain the field equations for each harmonic mode.
For the odd part which has the odd parity $(-1)^{\ell+1}$,
in the RW gauge in which $h_2=0$,
the following three equations (Z:C6a-c) are derived.
\begin{eqnarray}
\frac{\partial^2 h_0}{\partial r^2}
-\frac{\partial^2 h_1}{\partial t\partial r}
-\frac{2}{r}\frac{\partial h_1}{\partial t}
+\left[\frac{4M}{r^2}-\frac{\ell(\ell+1)}{r}\right]
\frac{h_0}{r-2M} &=&
\frac{8\pi}{\sqrt{\ell(\ell+1)/2}}
\frac{r^2}{r-2M}Q_{\ell m}^{(0)}\quad [CRTD]\,, \\
\frac{\partial^2 h_1}{\partial t^2}
-\frac{\partial^2 h_0}{\partial t \partial r}
+\frac{2}{r}\frac{\partial h_0}{\partial t}
+\frac{(\ell-1)(\ell+2)(r-2M)}{r^3}h_1 &=&
-\frac{8\pi i(r-2M)}{\sqrt{\ell(\ell+1)/2}}
Q_{\ell m}\quad [CRTD]\,, \\
\frac{\partial}{\partial r}
\left[\left(1-\frac{2M}{r}\right)h_1\right]
-\frac{r}{r-2M}\frac{\partial h_0}{\partial t} &=&
-\frac{8\pi ir^2}{\sqrt{\ell(\ell+1)(\ell-1)(\ell+2)/2}}
D_{\ell m}\quad [CRTD]\,.
\end{eqnarray}
Fo the even part which has the even parity $(-1)^{\ell}$,
we have seven equations (Z:C7a-g) in the RW gauge in which
$h_{0}^{(e)}=h_{1}^{(e)}=G=0$.
\begin{eqnarray}
\left(1-\frac{2M}{r}\right)^2\frac{\partial^2 K}{\partial r^2}
+\frac{1}{r}\left(1-\frac{2M}{r}\right)
\left(3-\frac{5M}{r}\right)\frac{\partial K}{\partial r}
& & \nonumber \\
-\frac{1}{r}\left(1-\frac{2M}{r}\right)^2
\frac{\partial H_2}{\partial r}
-\frac{1}{r^2}\left(1-\frac{2M}{r}\right)(H_2-K)
& & \nonumber \\
-\frac{\ell(\ell+1)}{2r^2}\left(1-\frac{2M}{r}\right)(H_2+K)
&=& -8\pi A_{\ell m}^{(0)}\quad [CRTD]\,, \\
\frac{\partial}{\partial t}\left[
\frac{\partial K}{\partial r}+\frac{1}{r}(K-H_2)
-\frac{M}{r(r-2M)}K \right]
& & \nonumber \\
-\frac{\ell(\ell+1)}{2r^2}H_1 &=&
-4\sqrt{2}\pi iA_{\ell m}^{(1)}\quad [CRTD]\,, \\
\left(\frac{r}{r-2M}\right)^2\frac{\partial^2 K}{\partial t^2}
-\frac{r-M}{r(r-2M)}\frac{\partial K}{\partial r}
-\frac{2}{r-2M}\frac{\partial H_1}{\partial t}
& & \nonumber \\
+\frac{1}{r}\frac{\partial H_0}{\partial r}
+\frac{1}{r(r-2M)}(H_2-K) & &
\nonumber \\
+\frac{\ell(\ell+1)}{2r(r-2M)}(K-H_0) &=& -8\pi A_{\ell m}
\quad [CRTD]\,, \\
\frac{\partial}{\partial r}\left[
\left(1-\frac{2M}{r}\right)H_1\right]
-\frac{\partial}{\partial t}(H_2+K) &=&
\frac{8\pi ir}{\sqrt{\ell(\ell+1)/2}}
B_{\ell m}^{(0)}\quad [CRTD]\,, \\
-\frac{\partial H_1}{\partial t}
+\left(1-\frac{2M}{r}\right)
\frac{\partial}{\partial r}(H_0-K)
+\frac{2M}{r^2}H_0
& & \nonumber \\
+\frac{1}{r}\left(1-\frac{M}{r}\right)(H_2-H_0) &=&
\frac{8\pi(r-2M)}{\sqrt{\ell(\ell+1)/2}}B_{\ell m}
\quad [CRTD]\,, \\
-\frac{r}{r-2M}\frac{\partial^2 K}{\partial t^2}
+\left(1-\frac{2M}{r}\right)\frac{\partial^2 K}{\partial r^2}
+\frac{2}{r}\left(1-\frac{M}{r}\right)
\frac{\partial K}{\partial r}
& & \nonumber \\
-\frac{r}{r-2M}\frac{\partial^2 H_2}{\partial t^2}
+2\frac{\partial^2 H_1}{\partial t\partial r}
-\left(1-\frac{2M}{r}\right)\frac{\partial^2 H_0}{\partial r^2}
& & \nonumber \\
+\frac{2(r-M)}{r(r-2M)}\frac{\partial H_1}{\partial t}
-\frac{1}{r}\left(1-\frac{M}{r}\right)
\frac{\partial H_2}{\partial r}
-\frac{r+M}{r^2}\frac{\partial H_0}{\partial r}
& & \nonumber \\
+\frac{\ell(\ell+1)}{2r^2}(H_0-H_2) &=&
8\sqrt{2}\pi G_{\ell m}^{(s)}\quad [CRTD]\,, \\
\frac{H_0-H_2}{2} &=&
\frac{8\pi r^2F_{\ell m}}{\sqrt{\ell(\ell+1)(\ell-1)(\ell+2)/2}}
\quad [CRTD]\,.
\end{eqnarray}

We now consider the Fourier transform of the above field equations.
The Fourier coefficients are defined, for example, as
\begin{eqnarray}
h_{0\ell m \omega}(r) = \int_{-\infty}^{+\infty}
dt \, h_{0\ell m}(t,r) e^{i\omega t} \,.
\end{eqnarray}
Then we derive the Regge-Wheeler-Zerilli equations
and construct the metric perturbation under the RW gauge condition.

For the odd part, a
 new radial function $R_{\ell m \omega}^{{\rm (odd)}}(r)$
is introduced, in terms of which
the two radial functions $h_{0\ell m \omega}$ and
$h_{1\ell m \omega}$
for the metric perturbation are expressed as
\begin{eqnarray}
h_{1\ell m \omega}&=&{r^2  \over r-2M}R_{\ell m \omega}^{{\rm (odd)}} \,,
\label{oddh1}\\
h_{0\ell m \omega}
&=&{i \over \omega}{d \over dr^*}(r R_{\ell m \omega}^{{\rm (odd)}})
-{8 \pi r(r-2M) \over
\omega [{1\over2}\ell(\ell+1)(\ell-1)(\ell+2)]^{1/2}}D_{\ell m\omega} \,.
\label{oddh0}
\end{eqnarray}
The new radial function satisfies
the Regge-Wheeler equation (Z:11),
\begin{eqnarray}
{d^2 R_{\ell m \omega}{}^{{\rm (odd)}} \over dr^{*2}}
&+&[\omega ^2 -V_{\ell}^{{\rm (odd)}}(r)]
R_{\ell m \omega}^{{\rm (odd)}}
\nonumber \\
&=&{8\pi i \over [{1\over2}l(\ell+1)(\ell-1)(\ell+2)]^{1/2}}{r-2M \over r^2}
\nonumber \\ &&
\times \left(-r^2{d \over dr}[(1-{2M \over r})D_{\ell m \omega}]
+(r-2M)[(\ell-1)(\ell+2)]^{1/2}Q_{\ell m \omega} \right)
\,[CRTD] \,,
\end{eqnarray}
where
$r^*=r+2M \log (r/2M-1)$ and
\begin{eqnarray}
V_{\ell}^{{\rm (odd)}}(r)
=\left(1-{2M \over r}\right)
\left({\ell(\ell+1) \over r^2}-{6M \over r^3}\right) \,.
\end{eqnarray}

For the even part, a new radial function
$R_{\ell m \omega}^{{\rm (even)}}(r)$ is introduced, in terms of which
the four radial functions (Z:13-16) are expressed as
\begin{eqnarray}
K_{\ell m \omega}
&=&{\lambda (\lambda +1)r^2+3\lambda Mr+6M^2 \over r^2(\lambda r+3M)}
R_{\ell m \omega}^{{\rm (even)}}
+{r-2M \over r}{d R_{\ell m \omega}^{{\rm (even)}} \over dr}
\nonumber \\
&& -{r(r-2M) \over \lambda r+3M}\tilde C_{1\ell m \omega}
+{i (r-2M)^2 \over r(\lambda r+3M)}\tilde C_{2\ell m \omega}
\quad [CRTD]\,,
\label{evenK}\\
H_{1\ell m \omega}&=&-i \omega
{\lambda r^2-3\lambda Mr-3M^2 \over (r-2M)(\lambda r+3M)}
R_{\ell m \omega}^{{\rm (even)}}
-i \omega r{d R_{\ell m \omega}^{{\rm (even)}} \over dr}
\nonumber \\
&& +{i \omega r^3 \over \lambda r+3M}\tilde C_{1\ell m \omega}
+{\omega r (r-2M) \over \lambda r+3M}\tilde C_{2\ell m \omega}
\quad [CRTD]\,,
\label{evenH1} \\
H_{0\ell m \omega}
&=&{\lambda r(r-2M)-\omega ^2 r^4 +M(r-3M) \over (r-2M)(\lambda r+3M)}
K_{\ell m \omega}
+{M(\lambda +1)-\omega ^2 r^3 \over i \omega r(\lambda r+3M)}H_{1\ell m \omega}
+\tilde B_{\ell m \omega} \quad [CRTD]
\,,
\label{evenH0}\\
H_{2\ell m \omega}
&=&H_{0\ell m \omega}-16 \pi r^2
[{1\over2}\ell(\ell+1)(\ell-1)(\ell+2)]^{-1/2}F_{\ell m \omega}
\quad [CRTD]\,,
\label{evenH2}
\end{eqnarray}
where we have introduced the symbol $\lambda$ for
\begin{eqnarray}
\lambda={1 \over 2}(\ell-1)(\ell+2) \,,
\end{eqnarray}
and the local source terms ((Z:17), (Z:20) and (Z:21)) by
\begin{eqnarray}
\tilde B_{\ell m \omega}
&=&{8\pi r^2(r-2M) \over \lambda r+3M}\{A_{\ell m \omega}
+[{1\over2}\ell(\ell+1)]^{-1/2}B_{\ell m \omega}\}
-{4\pi \sqrt{2} \over \lambda r +3M}
{Mr \over \omega}A_{\ell m \omega}^{(1)} \,, \\
\tilde C_{1\ell m \omega}&=&{8\pi \over \sqrt{2}\omega}A_{\ell m \omega}^{(1)}
+{1 \over r}\tilde B_{\ell m \omega}
-16\pi r [{1 \over 2}\ell(\ell+1)(\ell-1)(\ell+2)]^{-1/2}
F_{\ell m \omega} \quad [CRTD]\,,
\\
\tilde C_{2\ell m \omega}&=&-{8\pi r^2 \over i \omega}
{[{1\over2}\ell(\ell+1)]^{-1/2} \over r-2M}B_{\ell m \omega}{}^{(0)}
-{ir \over r-2M}\tilde B_{\ell m \omega}
\nonumber \\ &&
+{16\pi ir^3 \over r-2M}[{1 \over 2}\ell(\ell+1)(\ell-1)(\ell+2)]^{-1/2}
F_{\ell m \omega} \quad [CRTD]
\,.
\end{eqnarray}
We note that the above radial functions for the metric perturbation
have the local source terms
which have the $\delta$-function behavior at the particle location.
The new radial function obeys the wave equation,
\begin{eqnarray}
{d^2 R_{\ell m \omega}{}^{{\rm (even)}} \over dr^{*2}}
+\left[\omega ^2 -V_{\ell}^{{\rm (even)}}(r)\right]
R_{\ell m \omega}^{{\rm (even)}}
=S_{\ell m \omega} \,,
\label{eq:elederi}
\end{eqnarray}
where
\begin{eqnarray}
V_{\ell}^{{\rm (even)}}(r)=\left(1-{2M \over r}\right)
{2\lambda ^2(\lambda +1)r^3+6\lambda ^2 Mr^2+18\lambda M^2r+18M^3
\over r^3(\lambda r+3M)^2} \,,
\end{eqnarray}
and the source term is
\begin{eqnarray}
S_{\ell m \omega}&=&-i{r-2M \over r}{d \over dr}
\left[{(r-2M)^2 \over r(\lambda r+3M)}\left({ir^2 \over r-2M}
\tilde C_{1\ell m \omega}
+\tilde C_{2\ell m \omega} \right)\right]
\nonumber \\ &&
+i{(r-2M)^2 \over r(\lambda r+3M)^2}
\left[{\lambda (\lambda +1)r^2+3\lambda Mr+6M^2 \over r^2}
\tilde C_{2\ell m \omega}
+i{\lambda r^2-3\lambda Mr-3M^2 \over r-2M}
\tilde C_{1\ell m \omega}\right] \,.
\end{eqnarray}
The above equation (\ref{eq:elederi}) is called the Zerilli equation.
The Zerilli equation can be transformed to the Regge-Wheeler equation
by the Chandrasekhar transformation \cite{chandra}.
So we may focus only on the Regge-Wheeler equation if desired.
The Regge-Wheeler homogeneous solutions are discussed
in detail by Mano et al. \cite{Mano1}. Using their method,
one can construct the retarded Green function to
solve the inhomogeneous Regge-Wheeler equation.
Then the metric perturbation in the RW gauge is obtained from
Eqs.~(\ref{oddh1}) and (\ref{oddh0}) for the odd part and
from Eqs.~(\ref{evenK}-\ref{evenH2})  for the even part.

\end{appendix}



\begin{thebibliography}{99}

\bibitem{Mino:1996nk}
Y.~Mino, M.~Sasaki and T.~Tanaka,
Phys.\ Rev.\ D {\bf 55}, 3457 (1997)
[arXiv:gr-qc/9606018].

\bibitem{Quinn:1999kj}
T.~C.~Quinn and R.~M.~Wald,
Phys.\ Rev.\ D {\bf 60}, 064009 (1999)
[arXiv:gr-qc/9903014].

\bibitem{Detweiler:2002mi}
S.~Detweiler and B.~F.~Whiting,
arXiv:gr-qc/0202086.

\bibitem{Chrzanowski:wv}
P.~L.~Chrzanowski,
Phys.\ Rev.\ D {\bf 11}, 2042 (1975).

\bibitem{Regge:1957td}
T.~Regge and J. A. Wheeler,
Phys.\ Rev.\  {\bf 108}, 1063 (1957).

\bibitem{Zerilli:wd}
F.~J.~Zerilli,
Phys.\ Rev.\ D {\bf 2}, 2141 (1970).

\bibitem{Teukolsky:1973ha}
S.~A.~Teukolsky,
Astrophys.\ J.\  {\bf 185}, 635 (1973).

\bibitem{Mode-sum}
L.~Barack and A.~Ori,
Phys.\ Rev.\ D {\bf 61}, 061502 (2000)
[arXiv:gr-qc/9912010],
L.~Barack,
Phys.\ Rev.\ D {\bf 62}, 084027 (2000)
[arXiv:gr-qc/0005042].

\bibitem{Mino:2001mq}
Y.~Mino, H.~Nakano and M.~Sasaki,
arXiv:gr-qc/0111074.

\bibitem{Barack:2001gx}
L.~Barack, Y.~Mino, H.~Nakano, A.~Ori and M.~Sasaki,
Phys.\ Rev.\ Lett.\  {\bf 88}, 091101 (2002)
[arXiv:gr-qc/0111001].

\bibitem{Barack:2002mh}
L.~Barack and A.~Ori,
arXiv:gr-qc/0204093.

\bibitem{Barack:2001ph}
L.~Barack and A.~Ori,
Phys.\ Rev.\ D {\bf 64}, 124003 (2001)
[arXiv:gr-qc/0107056].

\bibitem{Barack:2002ku}
L.~Barack and C.~O.~Lousto,
arXiv:gr-qc/0205043.

\bibitem{Mano1}
S.~Mano, H.~Suzuki and E.~Takasugi,
Prog.\ Theor.\ Phys.\  {\bf 95}, 1079 (1996)
[arXiv:gr-qc/9603020],
S.~Mano, H.~Suzuki and E.~Takasugi,
Prog.\ Theor.\ Phys.\  {\bf 96}, 549 (1996)
[arXiv:gr-qc/9605057].

\bibitem{Mino} Y. Mino, talk given at the 5th Capra Ranch meeting,
http://cgwp.gravity.psu.edu/events/Capra5/slides/mino.ppt

\bibitem{SNS2} H. Nakano, N. Sago and M. Sasaki, in preparation.

\bibitem{chandra}
S. Chandrasekhar, Proc. R. Soc. (London) {\bf A343},
289 (1975); {\it Mathematical Theory of Black Holes\/}
(Oxford University Press, 1983), \\
M.~Sasaki and T.~Nakamura,
Phys.\ Lett.\ A {\bf 89}, 68 (1982).

\end{thebibliography}
\end{document}